\documentclass[a4paper,fleqn,usenatbib]{mnras}
\usepackage{newtxtext,newtxmath}
\usepackage[T1]{fontenc}
\usepackage{ae,aecompl}
\usepackage{graphics}
\usepackage{graphicx}
\usepackage{amsmath}	
\usepackage[flushleft]{threeparttable}
\usepackage{booktabs,caption}

\title[1E 1145.1-6141]{NuSTAR observation of X-ray pulsar 1E 1145.1-6141.}

\author[Ghising et. al.]{
Manoj Ghising,$^{1}$\thanks{manojghising26@gmail.com}
Mohammed Tobrej,$^{1}$\thanks{tabrez.md565@gmail.com}
Binay Rai,$^{1}$\thanks{binayrai21@gmail.com}
 Ruchi Tamang,$^{1}$\thanks{ruchitamang76@gmail.com}
\newauthor
Bikash Chandra Paul$^{1}$\thanks{bcpaul@associates.iucaa.in}
\\
$^{1}$Department of Physics, North Bengal University,Siliguri, Darjeeling, WB, 734013, India
\\
}

\pubyear{2021}

\begin{document}
\label{firstpage}
\pagerange{\pageref{firstpage}--\pageref{lastpage}}
\maketitle

\begin{abstract}
In this paper, we report on the hard X-ray observation of the X-ray pulsar 1E 1145.1-6141 performed with the Nuclear Spectroscopic Telescope Array mission (NuSTAR). The coherent pulsation of the source with a period of $\sim296.653\;\pm\;0.021\;s $ is detected. The source may be in the equilibrium phase, according to the most recent measurements of its pulse period. The pulse profile reveals a mild energy dependence and generally hints at a pencil-beam pattern. The pulse profile have evolved with time. The Pulse fraction is found to depend on energy with a fall in the value at $\sim 32\; keV$. The NuSTAR spectra can be approximated by a composite model with two continuum components, a blackbody emission, cut-off powerlaw, and a discrete component in the form of gaussian to account for the emission line of iron. The estimated absorbed flux of the source is $\sim6\times10^{-10}\;erg\;cm^{-2}\;s^{-1}$ which corresponds to a  luminosity of $\sim5\times 10^{36}\;erg\;s^{-1}$. Pulse phase-resolved spectroscopy were performed to understand the evolution of spectral parameters with pulse phase. The estimated blackbody radius is found to be consistent with the size of the theoretical prediction.

\end{abstract}


%
%
\section{Introduction}
X-ray binaries (XRBs) can be categorized based on their observational and physical characteristics. XRBs can be divided into three categories based on the mass of the companion star in the binary: low mass X-ray binaries (LMXBs), intermediate-mass X-ray binaries (IMXBs), and high mass X-ray binaries (HMXBs). Accreting HMXBs are known to be among the brightest X-ray sources in our Galaxy \citep{Nagase1989}. In these systems of binary star, both a neutron star and a massive main-sequence star rotate around the common center of mass in a wide and eccentric orbit \citep{Tauris2006}. The accretion of matter by a neutron star from the companion takes place through the capture of either stellar wind or Roche-lobe overflow. HMXBs with neutron star are categorized into Be/X-ray binaries and Supergiant X-ray binaries \citep{Reig2011}. It is now known that majority of the HMXBs are Be/X-ray binaries in which the mass-donor consists of a non-supergiant B or O spectral type star \citep{Reig2011}.

The \textit{Ariel} 5 observations in late 1978 discovered two X-ray pulsations from the source 4U 1145-619 (also known as 2S 1145-61) \citep{10}. Later on, \textit{Einstein} Observatory resolved this problem by identifying the other source to be 1E 1145.1-6141. The source 1E 1145.1-6141  is only 15$^{\prime}$ to the north of 2S 1145-619 \citep{9,10}. These two X-ray pulsars in Centaurus with close identical spin periods are also referred to as twin pulsars. The source 2S 1145-619 is the 292 s transient X-ray pulsar which is at a distance of 1.5 kpc.

The X-ray pulsar 1E 1145.1-6141 falls under the class of Supergiant HMXB. The source is associated with the B2 Iae supergiant companion V830 Cen and has a pulsation of $ \sim $ 297 s (\cite{7}; \cite{8}). The source is assumed to be at a distance of 8.5$\pm$ 1.5 kpc \citep{8}. With a typical X-ray flux of a few mcrab, or a luminosity of order $10^{36}\;erg\;s^{-1}$, the pulsar appears persistent and steady. A low luminosity suggests that the pulsar is almost certainly accreting from the wind of V830 Cen and is incompatible with Roche lobe overflow. The supergiant X-ray pulsar 1E 1145.1-6141 has an eccentricity of 0.20 and an orbital period of 14.4 days \citep{Ray_Chakrabarty2002}. On a $P_{spin}\;vs\;P_{orb} \; $ diagram, the system falls under the wind-fed supergiant XRBs \citep{Corbet1986,Water1989,Bildsten1997}. 
 
\section{Observations and Data Reduction}

\subsection{NuSTAR}
The data reduction for 1E 1145.1-6141 has been carried out using \textsc{heasoft} v6.29 and \textsc{caldb} version 1.0.2. NuSTAR operates in the energy range of (3-79) keV  and is the first hard X-ray focusing telescope. It consists of two identical X-ray telescope modules equipped with independent mirror systems. Each of the telescopes has its own focal plane modules A and B referred to as \textsc{fpma} \& \textsc{fpmb} consisting of a pixelated solid-state CdZnTe detector (\cite{1}). The detector CdZnTe of \textsc{fpma} \& \textsc{fpmb} provides spectral resolution of 400 eV (FWHM) at energy 10 keV. The presence of a unique multilayered coating of the grazing-incidence Wolter-I optics helps in providing X-ray imaging in the energy range (3-79) keV with an angular resolution of 18" (FWHM) \& 58" (HPD). 
The data extraction and screening has been done using \textsc{nustardas} software v 0.4.9. The mission-specific \textsc{nupipeline} was run for filtering the unfiltered clean event files.  Using \textsc{xselect} command and ds9\footnote{\url{https://sites.google.com/cfa.harvard.edu/saoimageds9}} application software, a circular region of  110" around the source center and of the same size away from the source were selected as the source and background region files respectively. The selected source and background regions were utilized for extracting the necessary light curves \& spectra by imposing the mission-specific \textsc{nuproducts}.  Next, light curves were corrected by subtracting the background contamination using ftool \textsc{lcmath} for both the instrument \textsc{fpma} \& \textsc{fpmb}. \textsc{barycentric} corrections were done using \textsc{ftool barycorr}. The NuSTAR observation with observation ID 30501002002 (see Table 1) has been considered for performing the required temporal and spectral analysis of the system.

\begin{table}
 \begin{center}
 \begin{tabular}{clllc}
    \hline
     \hline
   Observatory	& Date of observation &	OBs ID	&	Exposure	\\
	&		&	& (in ksec)	\\
\hline 
\hline					
NuSTAR	&	2019-07-23 & 30501002002	&	44.24	\\

     \hline
      \hline
  \end{tabular}
  \caption{NuSTAR observation details of the source 1E 1145.1-6141.}
  \end{center}
 \end{table}

\section{TIMING ANALYSIS}
\subsection{Pulse period estimation}

The pulse period of the source was determined  using the orbital corrected light curve from both the telescope modules \textsc{fpma} \& \textsc{fpmb} in the wide energy range (3-79) keV with a default newbin time of 10 s. The orbital parameters  for reducing the effect of orbital modulation were taken from the works of \cite{Ray_Chakrabarty2002}. Using the Fast Fourier Transform (\textsc{FFT}) technique  by imposing the command \textsc{powspec}, we estimated the pulse period of the source. We further employed the epoch-folding technique \citep{4,5} by using ftool \textsc{efsearch} to estimate the precise pulsation of the source using the determined value of the period from \textsc{FFT}. This method is based on the $\chi^{2}$ maximization technique. Finally, the best period value is determined which is 296.653\;$\pm$\;0.021 s. The uncertainty in the measurements of pulsations was estimated using the method described in \cite{6}. For this, we generated 1000 simulated light curves \& obtained the corresponding pulsations for each curve using the task \textsc{efsearch}. Finally, we computed the standard deviation \& standard error and hence established the precise pulsation \& the uncertainty involved in the measurement of the pulse period. 
\subsection{Energy-resolved Pulse Profiles}
\begin{figure*}

\begin{center}
\includegraphics[angle=0,scale=0.6]{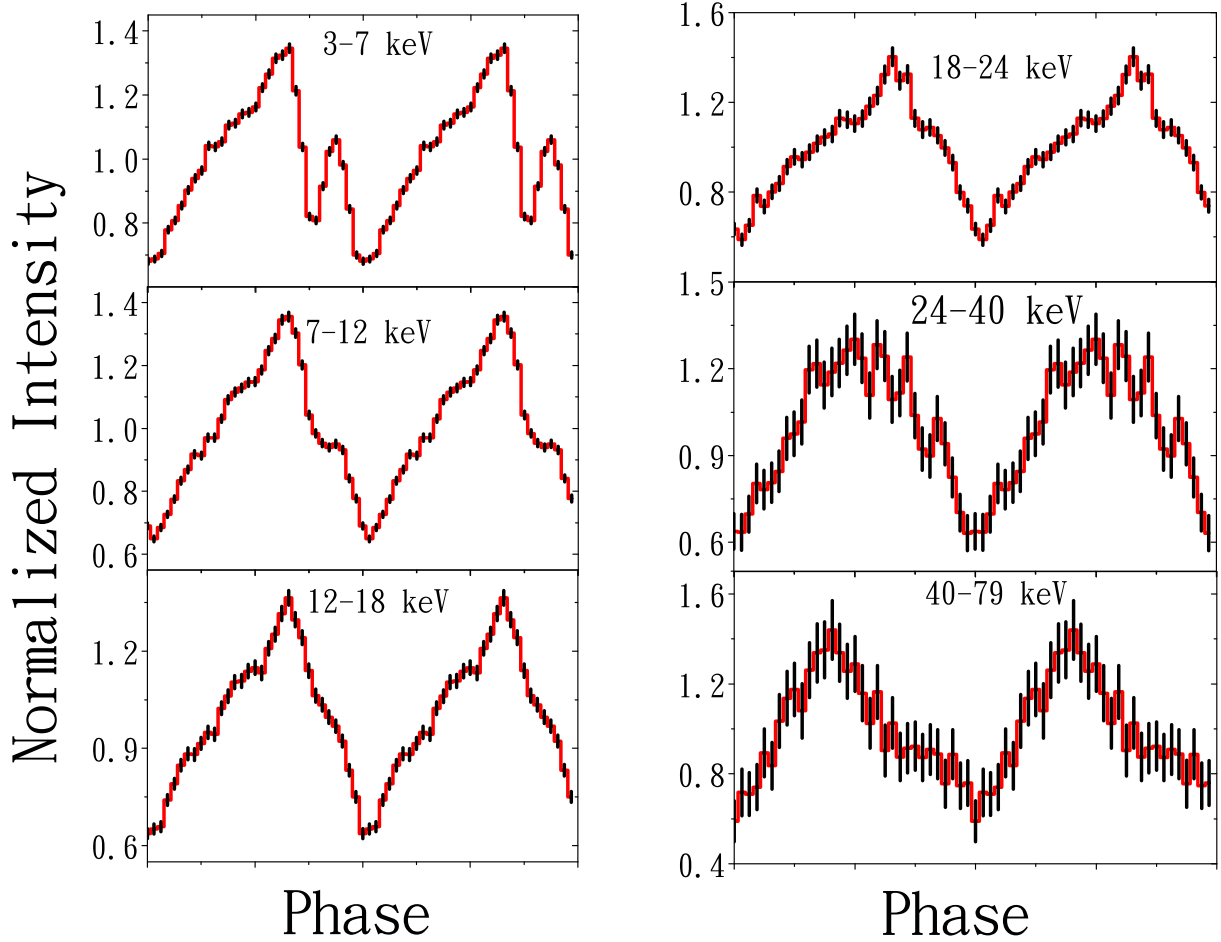}
\end{center}
\caption{Energy-resolved pulse profiles of 1E 1145.1-6141 for NuSTAR observation folded at the best period of 296.653$\pm$0.021 s}
\end{figure*}
In order to analyze energy-resolved pulse profiles (see Figure 1), we constructed light curves for the several energy bands, \textit{viz.} (3-7) keV, (7-12) keV, (12-18) keV, (18-24) keV, (24-40) keV, and (40-79) keV, by folding it at the determined value of the spin period. The pulse profile in the lowest energy band (3-7) keV is characterized by a broad single peak with an additional second peak in the phase interval 0.65-0.75. The second peak vanishes as the profile evolves with energy \textit{i.e} the pulse profile transforms to triangular shape in the harder energy ranges. The peak emission component shifts towards lower phases as the profile evolves with energy. 
\subsection{Pulse Fraction}
\begin{figure}

\begin{center}
\includegraphics[angle=0,scale=0.3]{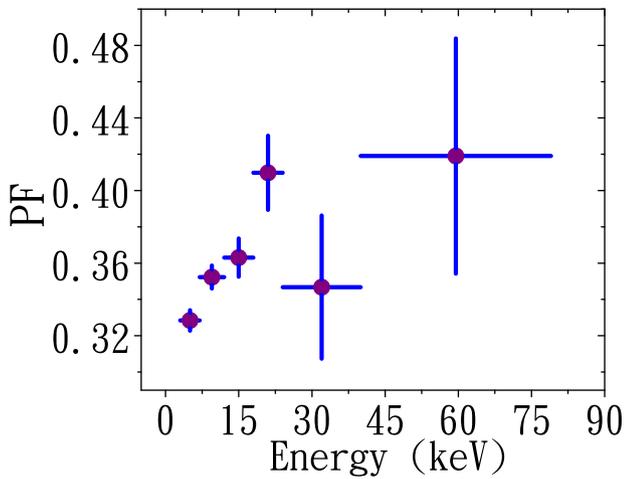}
\end{center}
\caption{Variation of Pulse Fraction (PF) with energy. The horizontal error bars indicate the energy range for which the pulsed fraction was calculated and the vertical error bars show errors
of corresponding measurements. The PF were obtained by dividing the (3-79) keV light curves into several energy bands.}
\end{figure}
The variation of Pulse Fraction (PF) with energy is represented in Figure 2. PF is defined as $\frac{P_{max}-P_{min}}{P_{max}+P_{min}}$ where $P_{max} \; \& \; P_{min}$ represents the
maximum and minimum intensities of the pulse profile. The PF initially increases with energy upto $\sim$20 keV followed by a drop in the value near $\sim$32 keV and above, the PF again increases with energy.

\section{SPECTRAL ANALYSIS} 

\subsection{Phase-average spectroscopy}
\begin{figure}

\begin{center}
\includegraphics[angle=270,scale=0.3]{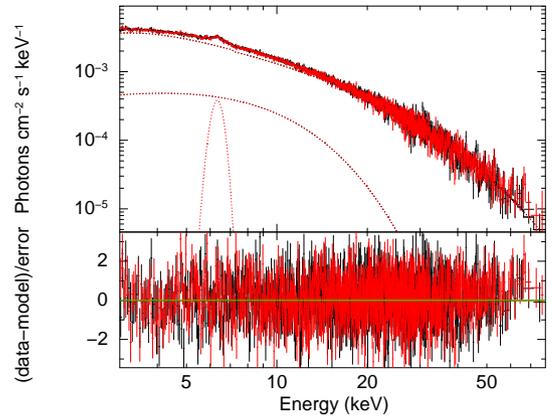}
\end{center}
\caption{Unfolded energy spectrum of 1E 1145.1-6141 and its approximation with the model combination \textsc{constant $\times$ tbabs $\times$ (cutoffpl+bbodyrad+gaussian)}. Red and black show data from the \textsc{fpma} \& \textsc{fpmb} telescopes of the NuSTAR observatory respectively. The phase-averaged spectrum does not reveal any characteristic absorption feature.}
\end{figure}

\begin{table*}
 \begin{center}
 \begin{tabular}{clllllllllllc}
    \hline
    \hline
Parameters	&	&	&	MODEL I	&	&	&	MODEL II	&	&	&	MODEL III	\\
\hline
\hline													
$C_{FPMA}$	&	&	&	1(fixed)	&	&	&	1(fixed)	&	&	&	1(fixed)	\\
$C_{FPMB}$	&	&	&	0.988$^{+0.003}_{-0.003}$	&	&	&	0.988$^{+0.003}_{-0.003}$	&	&	&	0.986$^{+0.003}_{-0.003}$	\\
$n_{H}\;(\times10^{22}\;cm^{-2})$	&	&	&	4.99$^{+0.60}_{-0.61}$	&	&	&	4.57$^{+0.89}_{-0.99}$	&	&	&	1.20 (fixed)	\\
COMPTT ($T_{o}$) (keV)	&	&	&	-	&	&	&	-	&	&	&	1.12$^{+0.03}_{-0.04}$	\\
COMPTT (kT) (keV)	&	&	&	-	&	&	&	-	&	&	&	9.98$^{+0.75}_{-0.59}$	\\
COMPTT ($\tau$) (keV)	&	&	&	-	&	&	&	-	&	&	&	3.47$^{+0.26}_{-0.27}$	\\
$\Gamma$	&	&	&	0.90$^{+0.07}_{-0.07}$	&	&	&	1.06$^{+0.09}_{-0.10}$	&	&	&	-	\\
$E_{CUT}$ (keV)	&	&	&	17.04$^{+1.19}_{-1.02}$	&	&	&	6.16$^{+0.43}_{-0.55}$	&	&	&	-	\\
$E_{fold}$ (keV)	&	&	&	-	&	&	&	19.24$^{1.95}_{-1.59}$	&	&	&	-	\\
$kT_{BBODYRAD}$ (keV)	&	&	&	2.61$^{+0.18}_{-0.19}$	&	&	&	3.13$^{+0.30}_{-0.29}$	&	&	&	3.47$^{+0.17}_{-0.16}$	\\
NORM$_{kT}$ 	&	&	&	0.106$^{+0.030}_{-0.021}$	&	&	&	0.045$^{+0.023}_{0-0.014}$	&	&	&	0.083$^{+0.015}_{-0.013}$	\\
Fe line (keV)	&	&	&	6.33$^{+0.04}_{-0.04}$	&	&	&	6.35$^{+0.04}_{-0.06}$	&	&	&	6.35$^{+0.04}_{-0.04}$	\\
$\sigma_{Fe}$ (keV)	&	&	&	0.28$^{+0.06}_{-0.05}$	&	&	&	0.19$^{+0.07}_{-0.08}$	&	&	&	0.28$^{+0.05}_{-0.05}$	\\
Flux($\times   10^{-10}\;erg\;cm^{-2}\;s^{-1}$)	&	&	&	6.01$^{+0.07}_{-0.08}$	&	&	&	6.03$^{+0.22}_{-0.05}$	&	&	&	6.01$^{+0.04}_{-0.05}$	\\
$\chi^{2}_{\nu}$&	&	&	0.97	&	&	&	0.96	&	&	&	0.97	\\
\hline
\hline
  \end{tabular}
  \caption{The above table highlights the best-fitting spectral parameters for the phase-averaged NuSTAR spectrum of 1E 1145.1-6141. The parameter $n_{H}$ represents neutral hydrogen column density, $\Gamma$ represents photon-index of \textsc{POWERLAW} model, $E_{CUT}$ and $E_{fold}$ are the cutoff energy and folded energy of the CUTOFFPL component, COMPTT ($T_{o}$), COMPTT (kT), COMPTT ($\tau$) are the soft comptonization temperature, plasma temperature and optical depth of the COMPTT component, $kT_{BBODYRAD}$ and NORM$_{kT}$ represents the blackbody temperature and norm of the BLACKBODY component, Fe  and $\sigma_{Fe}$ represents the iron line and its equivalent width of GAUSSIAN component. $\chi^{2}_{\nu}$ is the reduced $\chi^{2}$ for the spectral fit. All errors are quoted within 90 per cent confidence level.
  MODEL I: \textsc{constant$\times$tbabs$\times$(cutoffpl+bbodyrad+gaussian)} MODEL II: \textsc{constant $\times$ tbabs $\times$(highecut$\times$powerlaw+bbodyrad+gaussian)}     MODEL III: \textsc{constant $\times$ tbabs $\times$(comptt+bbodyrad+gaussian)}}
  \end{center}
 \end{table*}
 
The source 1E 1145.1-6141's X-ray spectra were fitted in the energy range of 3-79 keV. The data from NuSTAR were grouped using the spectral fitting tool \textsc{grppha} to ensure that there were at least 25 counts per spectral bin. The aforementioned energy range was used to perform the simultaneous spectral fitting of \textsc{fpma} \& \textsc{fpmb}. A model CONSTANT was included to assure the normalisation factor between the two modules, \textsc{fpma} \& \textsc{fpmb}. In the former instrument, the constant parameter was set to unity, whereas in the latter, we left it unset. The spectra's fitting revealed that the constant factor corresponding to \textsc{fpmb} was $0.980\pm0.003$. We introduced the TBABS component with the solar abundances from \cite{Wilms2000} in order to account for interstellar absorption, and we included the GAUSSIAN component in order to fit the emission lines of Fe.
 An additional BBODYRAD component was also required to approximate the spectra. 

Several phenomenological models were applied to fit the spectra of the source. In order to approximate the spectra of the source, we have considered  three model combinations \textsc{constant$\times$tbabs$\times$(cutoffpl+bbodyrad+gaussian)}, \textsc{constant $\times$ tbabs $\times$(highecut $\times$ powerlaw+bbodyrad+gaussian)} \& \textsc{constant $\times$ tbabs $\times$(comptt+bbodyrad+gaussian)} as shown in Table 2. It was observed that the applied model combinations approximates the spectra very well. However, the  neutral hydrogen column density ($n_{H}$) for the combination \textsc{constant $\times$ tbabs $\times$ comptt} were fixed at the Galactic value in the direction of the source 1.2$\;\times\; 10^{22}\;cm^{-2}$ \citep{23}. The unfolded energy spectrum for the model combination \textsc{constant $\times$ tbabs $\times$ (cutoffpl+bbodyrad+gaussian)} has been shown in Figure 3. Interestingly, it can be seen from Table 2 that the value of $n_{H}$ is relatively higher than the observed Galactic value for the two model combinations \textsc{constant $\times$ tbabs $\times$ cutoffpl} \& \textsc{constant $\times$ powerlaw $\times$ highecut.} Using the command \textsc{flux} in \textsc{xspec}, we approximated the source absorbed flux, which we found to be $\sim6\;\times\;10^{-10}\;erg\;cm^{-2}\;s^{-1}$  and its associated luminosity to be $\sim5\;\times\;10^{36}\;erg\;s^{-1}$ assuming a distance of 8.5$\pm$ 1.5 kpc \citep{8}.

\section{Phase-resolved spectroscopy}

\begin{figure*}

\begin{center}
\includegraphics[angle=0,scale=0.6]{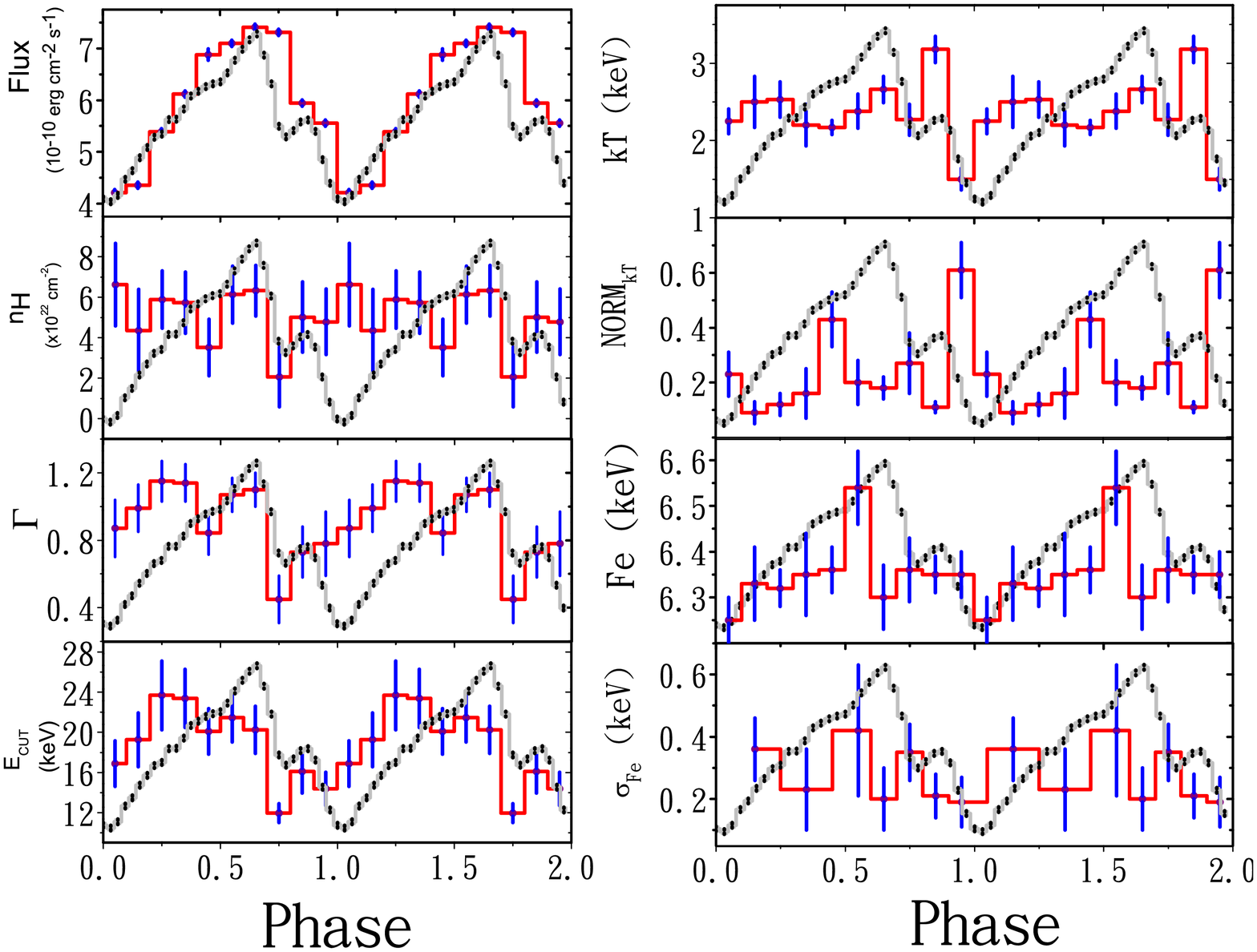}
\end{center}
\caption{Variation of spectral parameters with phase for the model combination \textsc{constant$\times$tbabs$\times$(cutoffpl+bbodyrad+gaussian)}. The parameter $n_{H}$ represents neutral hydrogen column density, $\Gamma$ and $E_{CUT}$ represents photon-index and the cutoff energy of the CUTOFFPL component, $kT_{BBODYRAD}$ and NORM$_{kT}$ represents the blackbody temperature and norm of the BLACKBODY component, Fe  and $\sigma_{Fe}$ represents the iron line and its equivalent width of GAUSSIAN component. The absorbed fluxes were estimated in the energy range (3-79) keV. The variation shown in the background in grey colour represents the (3-79) keV continuum pulse profile. Errors quoted for each parameter are within 90 \% confidence
interval. }
\end{figure*}

The phase-resolved spectroscopy was performed in an attempt to study the evolution of spectral parameters with the pulse phase of the pulsar. For this, the source pulse period was divided into ten equally distributed rotational phase bins. We then generated \textsc{good time interval (gti)} files based on the folding epoch and pulse period corresponding to each phase. By making use of GTI file we performed the task \textsc{nuproducts} for each phase and finally obtained the spectra files corresponding to ten pulse phases.
 We performed the analysis of all the ten phases in the energy range (3-79) keV with the model combination \textsc{constant$\times$tbabs$\times$(cutoffpl+bbodyrad+gaussian)} considered for average spectra. Figure 4 shows the evolution of the spectral parameters like Flux, $n_{H}$, $\Gamma$, $E_{CUT}$, kT, $NORM_{kT}$, Fe line and Fe width with the rotational phase. The zero points of all spectral parameters are considered at flux minimum for a representaive purpose so that we can visualize the evolution of spectral parameters that is present with the (3-79) keV pulse profile superimposed in grey color in the background (see Figure 4). It is obvious from Figure 4 that the spectral parameters evolve with a large amplitude over the pulse phase. The observed flux of the source correlates very well with the (3-79) keV pulse profile. The flux of the source varied between 24 \% and 30 \% \textit{i.e} (4.21-7.41) $\times\;10^{-10}\;erg\;cm^{-2}\;s^{-1}$ when compared with the average value. The photon-index ($\Gamma$) follows the main pulse profile with maximum value 1.15 in the phase interval 0.2-0.3 and minimum value 0.45 in the phase interval 0.7-0.8. The equivalent hydrogen column density ($n_{H}$) varied at all phases with a peak value of $6.63\;\times\;10^{22}\;cm^{-2}$ and a minimum value of $2.07\;\times\;10^{22}\;cm^{-2}$ in the phase intervals 0.7-0.8 and 0-0.1 respectively. The cutoff energy Though $(E_{CUT})$ has no physical meaning, but it is related the strength of the magnetic field \citep{Makishima1990}.  The variation of $(E_{CUT})$ is seen to depend at all phases with a maximum value of 23.67 keV in the phase interval 0.2-0.3 and a minimum value of 12 keV in the phase interval 0.7-0.8. The blackbody temperature (kT) of the emitting region varied between (1.50-2.66) keV in the phase interval 0.6-0.7 and 0.9-1.0 respectively. The blackbody emitting region \textit{i.e} the size of the emitting area is related to the norm of the BBODYRAD component and is found to vary between (0.05-0.15) km. The fluorescent line of iron varied between (6.25-6.54) keV while its equivalent width ($\sigma_{Fe}$) varied between (0.19-0.42) keV. However, the width at some phases were kept freezed at the value corresponding to the average spectral fit.

\section{Discussion and Conclusion}

In the paper, the temporal and spectral properties of the source 1E 1145.1-6141 have been examined by taking the data observed by NuSTAR with observation ID 30501002002. Previous work of \cite{Ray_Chakrabarty2002} on this source have shown the timing properties in detail, particularly, the orbital parameters. We detected the coherent pulsation of the source at 296.653$\;\pm\;0.021$ s. Past observations have reported the pulse period of the source as 296.572 ± 0.001 s and 296.695 ± 0.002 s \citep{33}, thereby, indicating that the source is undergoing through a phase of spin equilibrium \textit{i.e.} stable pulsar rotation.

The pulse profile of the source in (3-79) keV shown in grey color in the background (see Figure 4) is dominated by a  broad single peak with an additional second peak at phase $\sim$ 0.7. The additional secondary peak in the continuum profile is considered to be the reminiscent of the notch between the maximum and minimum pulse as reported by \cite{Grebenev1992} and \cite{Ray_Chakrabarty2002}. However, the additional emission component in the work of \cite{Ray_Chakrabarty2002} lies in the initial phase while in our present work it has shifted towards the initial phase. The variations of pulse profile with energy is nominal with the peak emission shifting towards the initial phase as observed by \cite{33}.

It has been noted by \cite{3} that pulse profile transforms from multi-peak at lower energy ranges to single-peak at higher energy ranges. Such features have been observed in many X-ray pulsars especially the bright ones like Her X-1, 4U 0115+63, 1A 0535+26 and V 0332+63. In particular, in the case of 1E 1145.1-6141, the additional emission component seen in the lowest energy range (3-7) keV disappears as the pulse profile evolves with energy and transforms into triangular shape in the higher energy ranges. 
The shape of the pulse profile is found to evolve with time when we compared that with the work of \cite{Ray_Chakrabarty2002} that infers a change in the accretion geometry.

Critical luminosity is defined as the luminosity below which the source accretes with pencil-beam geometry (known as sub-critical regime) and above which the source accretes with fan-beam geometry (super-critical regime). A pulsar's luminosity can affect the beaming pattern. Our timing analysis pulse profile suggest that the source may be accreting in the sub-critical domain. In the sub-critical regime, the beaming patttern is dominated by a pencil-beamed pattern where the accreted material penetrates the NS surface through collisions like nuclear (collisions with atmospheric proton) or coulomb (collisions with thermal electrons) \citep{Harding1994}. The emission in the case of a pencil-beamed pattern escapes from the top of the column \citep{Burnard_Aron_Klein1991}. In super-critical regime, the accreted material slows down towards the surface of a NS due to a radiation dominated shock created in the accretion column. As a result, the high radiation pressure stops the accretion material at a distance above the surface of a NS which in turn leads to X-ray emissions in the form of fan-beam shaped pattern \citep{Davidson1973}.

The presence of an absorption feature in the source can be inferred indirectly from the temporal analysis of the PF change with energy. Some X-ray pulsars have shown these characteristics in the past \citep{32,33,34,35,36}. An overall non-monotonic increase with energy can be seen in the PF variation with energy. The PF first rises with energy up to about 20 keV, which is characteristic of X-ray pulsars \citep{34}. Above 20 keV, a PF decrease is seen at about 32 keV, followed by an increase in PF at higher energies. A clue as to whether cyclotron lines are present in the spectra is provided by the source 1E 1145.1-6141's PF variation with energy. We sought confirmation as a result in both the average and the phase-resolved spectra. The source's average spectra do not exhibit any overt absorption features, such as cyclotron resonant scattering features in the energy range (3-79) keV. This led us to constrain the magnetic field of the source as either weaker than $\sim5\times10^{11}\;G$ or stronger than $\sim6\times10^{12}\;G$ when we consider the lower or the upper limit of the NuSTAR energy range (3-79) keV. Interestingly, recent studies have found that in some X-ray pulsars like GRO J2058+42 \citep{21} and Swift J1808.4-1754 \citep{22}, the CRSF feature although absent in the average continuum spectra, it is found to have feature phase-transient characteristic during the study of the pulse phase-resolved spectroscopy. In order to confirm the existence of the other absorption features, such as the cyclotron line or its higher harmonics, which may only be present at specific phase, we have used phase-resolved spectroscopy. However, we do not any such features using the model combinations used for average spectrum approximations.

The (3-79) keV spectrum was well approximated by an absorbed POWERLAW modified by a high energy CUTOFF, a phenomonological deconvolution, which are widely employed to model the spectra of accreting pulsars (e.g \cite{Coburn2002}). In addition, the spectrum  requires BBODYRAD and GAUSSIAN component to arrive at best spectral fit. Other components in the form of COMPTT and HIGHECUT were examined in place of CUTOFFPL which too approximated the spectra very well. The average flux of the source in the energy range (3-79) keV is $\sim6\;\times\;10^{-10}\;erg\;cm^{-2}\;s^{-1}$ and its corresponding luminosity is $5\;\times\;10^{36}\;erg\;s^{-1}$ assuming a distance of 8.5$\pm$ 1.5 kpc \citep{8}. Interestingly, it is observed that the best-fit value of $n_{H}$ for the two model components CUTOFFPL and COMPTT were 3-4 times higher than the Galactic value (1.2$\times\;10^{22}\;cm^{-2}$) in the direction of the source. The system may have a significant intrinsic absorption as one cause, or there may be some interstellar medium non-uniformities in the source's direction that make detection on typical maps challenging due to their poor angular resolution constraints.

Once the gas particles, or accreted matter, enter the magnetosphere, they can only move along the magnetic field lines. This causes the accretion flow to move in the direction of the magnetic poles, indicating that the accretion of matter is concentrated only in the polar caps, a small area on the surface of the NS. The magnetic field lines now adopt a form of column-shaped or funnel-shaped surface known as accretion column with polar caps at the base. The blackbody radius or emission area can be understood as the column's radius. A typical size of the polar caps lies between 0.1$R_{N}$ and 0.001$R_{N}$ \citep{Meszaros1992} , where $R_{N}$ denotes the radius of the NS. Assuming a radius of $\sim$10 km for a NS, the polar cap radius in our case lies close to $\sim$ 1 km. Thus the theoretical size of the polar caps and the size of the emission area that we determined are in good accord with \cite{Meszaros1992}.


 \section*{Data availability}
 
 The observational data used in this study can be accessed from the HEASARC data archive and is publicly  available for carrying out research work. 

\section{Acknowledgement}
This research work have used the NuSTAR data archived by the NASA High Energy
Astrophysics Science Archive Research Center (HEASARC) online
service maintained by the Goddard Space Flight Center. This work
has made use of the NuSTAR Data Analysis Software (NuSTARDAS)
jointly developed by the ASI Space Science Data Center (SSDC,
Italy) and the California Institute of Technology (Caltech, USA). We would like to express our gratitude to the anonymous reviewer for their insights and recommendations, which helped to shape the work as it stands now.









\bsp	
\label{lastpage}
\end{document}